\documentclass{asme2ej}
\pdfoutput=1
\usepackage{amssymb}
\usepackage{amsmath}
\usepackage{hyperref}
\usepackage{subfigure}% subcaptions for subfigures
\usepackage{subfigmat}% matrices of similar subfigures, aka small multiples
\usepackage{xcolor}
\usepackage{graphicx}
\hypersetup{colorlinks}
\usepackage{bm}

\graphicspath{{./}{./figs/}}

  \def\clap#1{\hbox to 0pt{\hss#1\hss}}

\providecommand{\mat}[1]{\bm{#1}}%
\renewcommand{\vec}[1]{\mathbf{#1}}

\providecommand{\mF}{\ensuremath{\mat{F}}}

\providecommand{\mK}{\ensuremath{\mat{K}}}

\providecommand{\mW}{\ensuremath{\mat{W}}}

\providecommand{\ve}{\ensuremath{\vec{e}}}

\providecommand{\vq}{\ensuremath{\vec{q}}}

\providecommand{\vt}{\ensuremath{\vec{t}}}

\newcommand{\sS}{\mathcal{S}}

\newcommand{\bmat}[1]{\begin{bmatrix}#1\end{bmatrix}}

\title{Density-Matching for Turbomachinery Optimization Under Uncertainty}

\author{Pranay Seshadri\thanks{Address all correspondence to ps583@cam.ac.uk. Copyright  (c) 2015 by Rolls-Royce plc.}
    \affiliation{
	PhD Research Student,\\
	Engineering Design Centre,\\
	Department of Engineering,\\
	University of Cambridge\\
	Cambridge, CB2 1PZ\\
    Email: ps583@cam.ac.uk
    }	
}

\author{Geoffrey T. Parks
    \affiliation{Reader,\\
    Engineering Design Centre,\\
    Department of Engineering,\\
    University of Cambridge\\
    Cambridge, CB2 1PZ\\
        Email: gtp10@cam.ac.uk
    }
}

\author{Shahrokh Shahpar
    \affiliation{Rolls-Royce Associate Fellow,\\
        CFD Methods, Rolls-Royce plc.\\
        Derby, DE24 8BJ\\
        Email: shahrokh.shahpar@rolls-royce.com
    }
}

\begin{document}

\maketitle    

%%%%%%%%%%%%%%%%%%%%%%%%%%%%%%%%%%%%%%%%%%%%%%%%%%%%%%%%%%%%%%%%%%%%%%
\begin{abstract}
{\it 
A monotonic, non-kernel density variant of the density-matching technique for optimization under uncertainty is developed. The approach is suited for turbomachinery problems which, by and large, tend to exhibit monotonic variations in the circumferentially and radially mass-averaged quantities---such as pressure ratio, efficiency and capacity---with common aleatory turbomachinery uncertainties. The method is successfully applied to de-sensitize the effect of an uncertainty in rear-seal leakage flows on the fan stage of a modern jet engine. 
}
\end{abstract}

%%\footnotetext{Copyright \copyright{} 2015 by Rolls-Royce plc.}
\section{Introduction}
\label{sec:Intro}
Aero-engine manufacturers are becoming increasingly reliant on computational predictions of compressor and turbine flows to guide their design decisions. As this trend will continue in the years to come, there will be a growing emphasis on leveraging these methods within optimization routines. In tandem, there will also be a demand for the application of uncertainty quantification (UQ) methods, which seek to compute error bars, probability density functions (pdfs) and probabilities of failure for a given compressor or turbine design. Computational simulations of turbomachinery flows can be used to yield these statistics in an efficient manner. Central to this endeavor is the identification of operational uncertainties and their variations. These may be of the form of operational variations in the tip gap, blade and endwall manufacturing deviations and leakage flows between rotating and stationary components---to name but a few. 

Finally, to close the loop for redesign, this statistical data needs to be fed to an optimizer in a suitable manner. Here one would like to depart from expensive multi-objective methods and opt for computationally tractable alternatives. A framework that can leverage adjoint-based design sensitivities is also extremely favorable. These perspectives give rise to the main focus area of this research. In this paper, a monotonic variant of a probability density-matching technique for optimization under uncertainty (OUU) is presented. A brief background on density-matching and its advantages over existing robust and reliability-based design optimization techniques is given in Section 2. In Section 3, the mathematics behind monotonic density-matching and the motivation for this technique is presented. Finally, in Section 4, the method is successfully applied to a real turbomachinery problem: fan sub-system design under rear-seal leakage uncertainty. 

\section{Literature review}
\label{sec:lit}
A brief literature review of OUU methods is provided below.

\subsection{Robust design optimization (RDO)}
Robust design methodologies seek to optimize the mean performance metric while minimizing its variance \cite{Messac}. Typically RDO lends itself to a multi-objective optimization strategy where there is a trade-off between these two objectives. This approach yields a Pareto set of feasible design solutions from which the designer will select a single design that offers a reasonable trade-off between the two objectives. By and large, multi-objective genetic algorithms are the optimizer of choice for these problems. Aerospace applications range from the design of compressor blades \cite{Seshadri_ASME} to entire compression systems \cite{Ghisu}, and airfoils \cite{Gianluca_windmill}. Most of the examples cited use low-fidelity computational simulations, where individual design simulations are obtained within minutes. This approach may not be viable for industrial design cases, where complex multi-physics simulations may take several hours or days. An alternative single-objective strategy for obtaining the Pareto front is scalarization, which minimizes a weighted sum of the two objectives. Typically multiple instances of the scalarization are required, each with a different set of weights, to obtain points on the Pareto front \cite{Allen}. One of the weaknesses of this approach is its inability to obtain parts of the Pareto set that are non-convex \cite{DasDennis}. 

\subsection{Reliability-based design optimization (RBDO)}
RBDO methods seek to minimize a cost function for an engineered system, while making allowances for a specific risk and target reliability under various sources of uncertainty \cite{Frangopol}. This involves the evaluation of reliability metrics that need to be computed within the optimization framework \cite{Youn}. RBDO can be formulated as a single- or multi-objective problem with reliability constraints that require tail statistics. Practical applications of RBDO include the optimization of a transonic compressor \cite{Lian}, an airfoil with nonlinear aeroelasticity \cite{Missoum}, structures \cite{Allen} and vehicle crash worthiness \cite{Carcrash}. 

\subsection{Density-Matching}
The RDO and RBDO methods emphasize the roles of low-order moments and tail probabilities, respectively, in the design optimization process when uncertainty is present in operating conditions. In \cite{Seshadri}, an alternative single-objective RDO / RDBO strategy---in which no moment definitions and no scalarization is required---was proposed. The approach takes into account the full pdf of the response with respect to the uncertainties. 

Termed density-matching, this method addresses the situation where a designer has expressed the desired performance of the system as a pdf, called the \emph{target pdf}. The objective in density-matching is to seek a point in the design space whose response pdf matches the given target as closely as possible. Using kernel density estimates of the pdf, density-matching is formulated as a continuous optimization problem suited for existing optimization software. Details of the approach are elaborated upon below.

Define a function $f=f(s,\omega)$ that represents the response of a physical model with design variables $s\in\mathcal{S}\subseteq\mathbb{R}^{n}$ and uncertain variables $\omega\in\Omega$. One can think of $f$ as the quantity of interest (qoi) to the designer. The space $\sS$ denotes the constraints on the design variables. For simplicity, assume that $f$ is scalar-valued, $f\in\mathcal{F}\subseteq\mathbb{R}$, though this can be generalized. We also assume that $f$ is continuous in both $s$ and $\omega$. The uncertain inputs $\omega$ are a vector of random variables (or in some cases a single variable) defined on a complete probability space $(\Omega,\mathcal{F},P)$ with a pdf $p=p(\omega)$ corresponding to the measure $P$. For a fixed  $s\in\mathcal{S}$, let $q_{s}:\mathcal{F}\rightarrow\mathbb{R}_{+}$  be a probability density function of $f(s,\omega)$. The shape of $q_s$ will be different for different values of $s$. We assume that we are given a pdf that describes the desired behavior of the model, including random variations. This pdf is the \emph{target pdf}, and it is provided by the designer.  For example, we may be given the mean and the standard deviation of a normal distribution, which respectively describe the desired average response and the permissible fluctuations. Denote the target pdf by $t:\mathbb{R}\rightarrow\mathbb{R}_+$.

The distance between the target pdf and the \emph{current design} pdf is given by the (squared) $L_2$-norm,
\begin{equation}
\label{eq:obj}
d(t,q_s) \;=\; \int_{-\infty}^{\infty} (t(f) - q_s(f))^2\,df,
\end{equation}
under the assumption that $t$ and $q_s$ are square-integrable. Moreover, it is also assumed that $f(s,\omega)$ is bounded for all $s$ and $\omega$, $f_\ell \leq f(s,\omega)\leq f_u$. An $N$-point numerical integration rule on the interval $[f_\ell,f_u]$ with points $\bar{f}_i$ and weights $w_i$ with $i=1,\dots,N$ is then selected to discretize the distance
\begin{equation}
\label{eq:aopt}
d(t,q_s) \;\approx\; \tilde{d}(t,q_s) 
\;=\; \sum_{i=1}^N (t(\bar{f}_i) - q_s(\bar{f}_i))^2\,w_i
\;=\; (\vt-\vq_s)^T\mW(\vt-\vq_s),
\end{equation}
where
\begin{equation}
\mW=\bmat{w_1 & & \\ & \ddots & \\ & & w_N},\quad
\vt=\bmat{t(\bar{f}_1)\\ \vdots \\ t(\bar{f}_N)},\quad
\vq_s=\bmat{q_s(\bar{f}_1)\\ \vdots \\ q_s(\bar{f}_N)}.
\end{equation}
For a fixed $s$, the density $q_s$ is, in general, not a known function of $f$ and must be estimated. To ensure that the objective function remains differentiable with respect to $s$, a kernel density estimate of $q_s$ can be used. This is done by choosing a set of $M$ points $\omega_j\in\Omega$ and defining the functions $f_j(s)=f(s,\omega_j)$ for $j=1,\dots,M$. Then, for a bandwidth parameter $h$ and a kernel $K=K_h$, $q_s$ can be approximated by
\begin{equation}
\label{eq:kde}
q_s(f) \;\approx\; \tilde{q}_s(f) \;=\; \frac{1}{M} \sum_{j=1}^M K(f-f_j(s)).
\end{equation}
Here the vector $\vq_s$ can be approximated by $\vq_s \approx \tilde{\vq}_s = \mK\ve$ where $\mK\in\mathbb{R}^{N\times M}$, $\mK_{ij}=\frac{1}{M} K(\bar{f}_i-f_j(s))$, and $\ve$ is an $M$-vector of ones. For computation, $\vq_s$ can be replaced by $\tilde{\vq}_s$ in the approximate objective function $\tilde{d}$ in Eq.~\eqref{eq:aopt}. This form lends itself to a gradient computation $\tilde{d}$ with respect to the design variables $s$. Now define the $M\times n$ matrix $\mF'$ by
\begin{equation}
\mF' \;=\; 
\bmat{
\frac{\partial f_1}{\partial s_1} & \cdots & \frac{\partial f_1}{\partial s_n}\\
\vdots & \ddots & \vdots\\
\frac{\partial f_M}{\partial s_1} & \cdots & \frac{\partial f_M}{\partial s_n}\\
} \qquad
\mK'_{ij} = \frac{1}{M} K'(\bar{f}_i-f_j(s)),
\end{equation}
where $n$ is the number of design variables. Oriented as a row vector, the gradient of the objective function can then be written as
\begin{equation}
\nabla_s \tilde{d} \;=\; 
2\,\left(\vt-\mK\ve\right)^T\,\mW\,\mK'\,\mF'.
\label{equation_gradients}
\end{equation}
This expression puts two restrictions on the types of problems that may be considered. First, a differentiable kernel $K$ in the kernel density estimator of $q_s$ in Eq.~\eqref{eq:kde} must be used. Second, only functions $f(s,\omega)$ that are differentiable with respect to the parameters $s$ can be used with the gradient formulation. In \cite{Seshadri}, the authors apply the above method to the design of an airfoil under inlet uncertainty with Gaussian kernels and adjoint sensitivities. Another point on this formulation is in order. For expensive simulations, it may be more efficient to use a response surface when approximating the pdf $q_s$. For instance in Seshadri et al., the authors use a 5th degree polynomial least squares. However, the number of samples used for the response surface is determined by the relationship between $\partial_f/\partial_{s_{k}}$, where $k=1,\ldots,n$ and the uncertainty and not $f$ and the uncertainty. This adds to the cost of the number of samples. In the next section, a monotonic variant of the same approach is presented which requires only two gradient evaluations, regardless of the relationship between $\partial_f/\partial_{s_{k}}$. 

\section{Monotonic density-matching}
\label{sec:montonic}
In this section, the derived distribution theorem (see Chapter 4 of \cite{MIT}) is used within the density-matching design framework to replace kernel density estimation with the true pdf. The objective in this section is to derive the density-matching formulation, for both distance and gradient computations---in the absence of kernel density estimates.

\subsection{Motivation}
Requiring monotonicity in the uncertainty-to-qoi relation severely restricts the types of problems that may be considered for OUU. The rationale for imposing this restriction is based on the nature of uncertainties in the turbomachinery environment and their impact on global qois such as pressure ratio, efficiency and capacity - as these usually form the objective functions and constraints for modern industrial turbomachinery design optimization. It should be noted that these optimizations are usually carried out at or close to peak efficiency and are thus far away from the flow-field nonlinearities associated with stall or choke. These metrics are derived from circumferentially and radially mass-averaged flow quantities. As a result, even though uncertainties may locally introduce nonlinear perturbations in the pressure or density, when averaged out, the effect of these uncertainties on the qois remains relatively linear. An overview of existing UQ techniques applied to both compressor and turbine flows finds that these metrics usually vary linearly as a function of uncertainties in tip gaps \cite{Seshadri_ASME, Montomoli_tip}, leakage flows\cite{Seshadri_LEAK}, inlet pressure perturbations \cite{Seshadri_old}. In fact, for the case study considered in Section 4, the effect of the uncertainty on efficiency is also found to be linear (see Figure~\ref{uncons_workflow}(a)). Here the variation in the rear seal leakage mass-flow rate---the uncertainty---is plotted against the stage efficiency---the qoi---for two different designs of a modern fan stage. 

\subsection{Monotonic derivation}
Without loss of generality, all computational uncertainties are characterized by $\mathcal{U}$ and the corresponding output quantities of interest (qoi) by $\mathcal{Q}$. The derived distribution theorem states that the pdf of a qoi can be defined as a function of the pdf of the inputs and the surrogate model:
%\begin{table}
% \begin{center}
%   \caption{Nomenclature}
%   \label{LADProb}
%   \begin{tabular}{ll}
%%%   \toprule
%    Property & Definition \\
%    \hline
%    $s$ & vector of design variables \\
%    $\mathcal{Q}\left(s\right) = f(\mathcal{U}\left(s\right))$ & continuous output qoi as a function of the input uncertainties  \\  
%    $\mathcal{\bar{Q}}_{i}\left(s\right)$ & discrete output qoi from simulation $i$ \\
%    $\mathcal{\bar{U}}_{i}\left(s\right)$ & discrete input uncertainty in simulation $i$ \\
%    $p\left(\cdot\right)$ & pdf of input uncertainty  \\
%    $r\left(\cdot\right)$ & pdf of output qoi  \\
%    $\left(\cdot\right)_U$ & upper bound of support \\ 
%    $\left(\cdot\right)_L$ & lower bound of support \\ 
%    $N$ & number of quadrature points for numerical integration \\ 
%    $k$ & number of design parameters \\ 
%%%   \toprule
%   \end{tabular}
% \end{center}
%\end{table}
\begin{align}
	r\left(\mathcal{Q}\right) = 
	\begin{cases}
		p\left(g^{-1}\right)\left|\frac{dg^{-1}}{d\mathcal{Q}}\right| & if \; \; \mathcal{U}_{L}\leq g^{-1}\left(\mathcal{Q}\right)\leq\mathcal{U}_{U} \\
		0 & \text{otherwise}.
	\end{cases}
	\label{lad_form}
\end{align}
Here $g$ is a surrogate model that must be differentiable, invertible and monotonic. In the general case $g$ can be written as $g=g\left(\mathcal{Q},\mathcal{\bar{Q}}_{1},\ldots\mathcal{\bar{Q}}_{l}\right)$, with the condition that $g^{-1}$ exists. The function $g$ approximates the continuous response $\mathcal{Q}$ using information from the outputs of discrete simulations $\mathcal{\bar{Q}}_{i}$. 
Using Eq.~\eqref{lad_form}, the density-matching design problem can be stated as minimizing the distance,
\begin{equation}
d(t,q_s) \;=\; \int_{-\infty}^{\infty} (t(\mathcal{Q}) - r(\mathcal{Q}))^2\,d\mathcal{Q},
\end{equation}
where $t(\mathcal{Q})$ is the familiar target distribution, to be selected by the designer.

\subsection{Monotonic density-matching}
The motivation for deriving monotonic density-matching with a linear surrogate is to develop the computational framework necessary for pursuing OUU with only 2 CFD runs per design. To obtain this surrogate, two simulations corresponding to two states, $(\mathcal{\bar{Q}}_1, \mathcal{\bar{U}}_1)$ and $(\mathcal{\bar{Q}}_2, \mathcal{\bar{U}}_2)$, are evaluated. This yields a more precise definition of $g$ as $g\left(\mathcal{Q},\mathcal{\bar{Q}}_{1},\mathcal{\bar{Q}}_{2}\right)=a\mathcal{U}+b,$ where $a=a\left(\mathcal{U},s\right)=\left(\mathcal{\bar{Q}}_{2}\left(s\right)-\mathcal{\bar{Q}}_{1}\left(s\right)\right)/\left(\mathcal{\bar{U}}_{2}\left(s\right)-\mathcal{\bar{U}}_{1}\left(s\right)\right)$, where the shift $b$ can be expressed as:
\begin{equation}
b=b\left(\mathcal{U},s\right)=\mathcal{\bar{Q}}_{2}\left(s\right)-a\mathcal{\bar{U}}_{2}\left(s\right)=\mathcal{\bar{Q}}_{2}\left(s\right)-\left(\frac{\mathcal{\bar{Q}}_{2}\left(s\right)-\mathcal{\bar{Q}}_{1}\left(s\right)}{\mathcal{\bar{U}}_{2}\left(s\right)-\mathcal{\bar{U}}_{1}\left(s\right)}\right)\mathcal{\bar{U}}_{2}\left(s\right)
\end{equation}
\begin{equation}
b\left(\mathcal{U},s\right)=\left(\frac{\mathcal{\bar{U}}_{1}\left(s\right)\mathcal{\bar{Q}}_{2}\left(s\right)-\mathcal{\bar{U}}_{2}\left(s\right)\mathcal{\bar{Q}}_{1}\left(s\right)}{\mathcal{\bar{U}}_{2}\left(s\right)-\mathcal{\bar{U}}_{1}\left(s\right)}\right).
\end{equation}
For consistency one assumes that $\mathcal{\bar{Q}}_{1}\left(s\right)\geq\mathcal{\bar{Q}}_{2}\left(s\right)$
where the qoi deteriorates with increasing uncertainty, that is $\mathcal{\bar{U}}_{1}\left(s\right)<\mathcal{\bar{U}}_{2}\left(s\right)$.
With adjoint sensitivities, the partial derivatives of the slope, $a$ can be computed:
\begin{equation}
\frac{\partial a}{\partial s_{k}}=\frac{1}{\mathcal{\bar{U}}_{2}\left(s\right)-\mathcal{\bar{U}}_{1}\left(s\right)}\cdot\left(\frac{\partial\mathcal{\bar{Q}}_{2}\left(s\right)}{\partial s_{k}}-\frac{\partial\mathcal{\bar{Q}}_{1}\left(s\right)}{\partial s_{k}}\right)
\end{equation}
where $\partial\mathcal{\bar{Q}}_{i}/\partial s_{k}$ is the set of adjoint sensitivities at the uncertainty state $\mathcal{\bar{U}}_{i}$. As before, $s_{k}$ refers to the set of design parameters. In a similar manner, the partial derivative of the shift can be computed as
\begin{equation}
\frac{\partial b}{\partial s_{k}}=\frac{1}{\mathcal{\bar{U}}_{2}\left(s\right)-\mathcal{\bar{U}}_{1}\left(s\right)}\cdot\left(\mathcal{\bar{U}}_{1}\left(s\right)\frac{\partial\mathcal{\bar{Q}}_{2}\left(s\right)}{\partial s_{k}}-\mathcal{\bar{U}}_{2}\left(s\right)\frac{\partial\mathcal{\bar{Q}}_{1}\left(s\right)}{\partial s_{k}}\right).
\end{equation}
Finally, for reasons that will become clear shortly, the partial derivative for $v=\left(y-b\right)/a$ can be easily written using the quotient rule
\begin{equation}
\frac{\partial v}{\partial s_{k}}=\frac{\left(-\partial b/\partial s_{k}\right)a-\left(\mathcal{Q}-b\right)\partial a/\partial s_{k}}{a^{2}}
\end{equation}
With these preliminaries, the monotonic density-matching problem can now be stated as 
\begin{equation}
minimize \;\; d=\left(\mathbf{t}-\mathbf{r}\right)^{T}\mathbf{W}\left(\mathbf{t}-\mathbf{r}\right)
\end{equation}
where $\mathbf{r}$ is the pdf of the current design:
\begin{equation}
\mathbf{r}=\left[\begin{array}{c}
r_{1}\\
\vdots\\
r_{N}
\end{array}\right],r_{i}=\frac{1}{\left|a\right|}p\left(\frac{\mathcal{Q}_{i}-b}{a}\right);i=1,\ldots,N.
\end{equation}
In the same manner as earlier, the gradient of the objective function ($d$) is written as $\nabla_{s}d=2\left(\mathbf{t}-\mathbf{r}\right)^{T}\mathbf{W}\mathbf{D}$, where 
\begin{equation}
\mathbf{D}=\left[\begin{array}{ccc}
\partial r_{1}/\partial s_{1} & \ldots & \partial r_{1}/\partial s_{k}\\
\vdots & \ddots & \vdots\\
\partial r_{N}/\partial s_{1} & \cdots & \partial r_{N}/\partial s_{k}
\end{array}\right]
\end{equation}
is a $k\times N$ matrix, where each entry is 
\begin{equation}
\frac{\partial r_{i}}{\partial s_{k}}=-\frac{1}{a^{2}}\frac{\partial a}{\partial s_{k}}r\left(\frac{\mathcal{Q}_{i}-b}{a}\right)+\frac{1}{a}r'\left(\frac{\mathcal{Q}_{i}-b}{a}\right)\frac{\partial v}{\partial s_{k}}.
\end{equation}

\subsection{Analytical example}
\label{sec:example}
\begin{figure}[t!]
\centering
\includegraphics[natwidth=930,natheight=396,width=13cm]{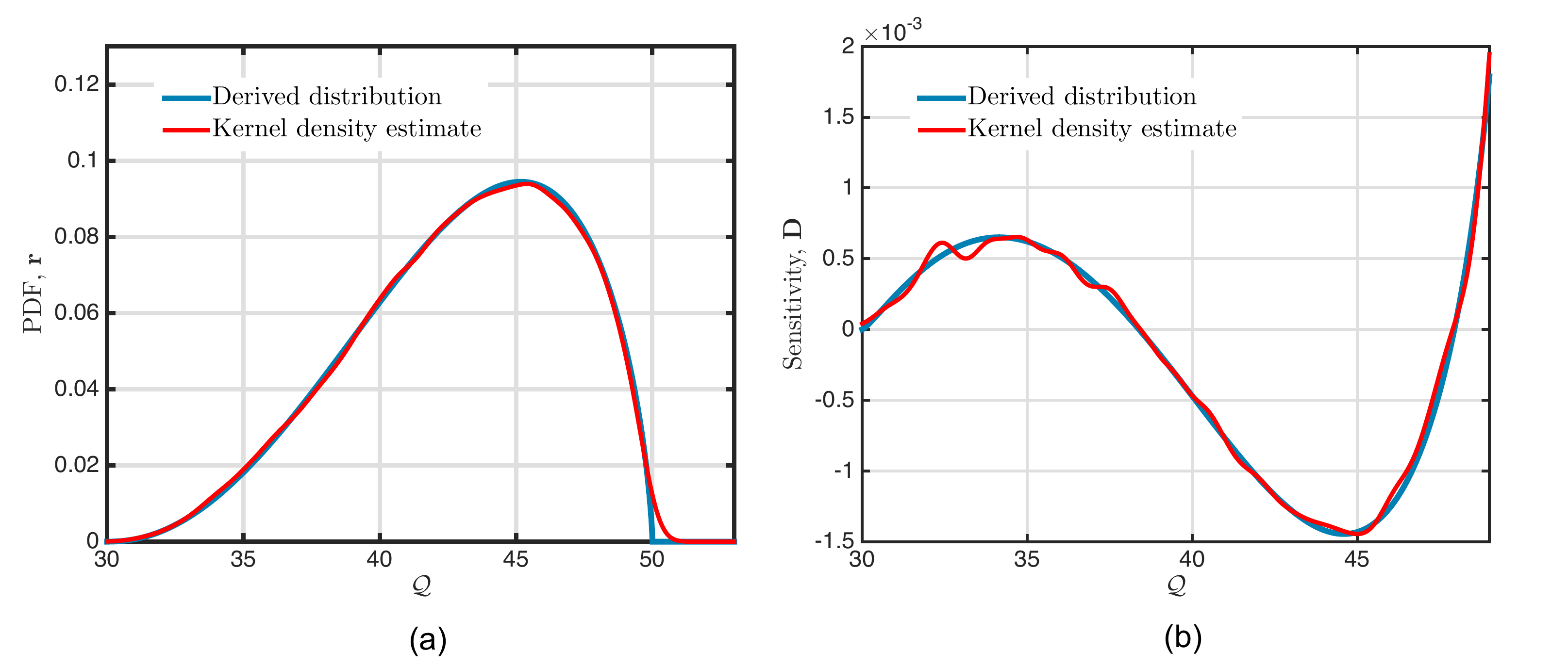}
\caption{Comparison of the monotonic formulation (shown in blue) with kernel density estimation (shown in red) for: (a) $\mathbf{r}$; (b) $\mathbf{D}$.}
\label{num_example}
\end{figure}

To clarify the above, consider the following linear model, $\mathcal{Q}=200\mathcal{U}+10$, where $\mathcal{U}$ is a beta $\left(1.7,3.2\right)$ random variable bounded by $\mathcal{U}\in\left[0.1,0.2\right]$. Assume that this linear model is obtained from two computational simulations with a single design parameter where 0.1 and -0.2 are the values of $\partial\mathcal{Q}_{1}/\partial s$ and $\partial\mathcal{Q}_{2}/\partial s$ respectively. From this data both the pdf of $\mathcal{Q}$, $\mathbf{r}$ and its sensitivity, $\mathbf{D}$, can be computed. Using the formulations outlined earlier, both the pdf and its sensitivity are plotted in Figure~\ref{num_example}. The kernel density estimation, which is shown here as a basis for comparison, was generated using $10^5$ random samples. Both the kernel estimate and the derived distribution plots were computed on 2000 equidistant quadrature points using a trapezoidal integration rule. As stated by Scott \cite{Scott}, the derivative of the kernel density estimate amplifies the oscillations within the kernel density estimate. While the kernel density estimate captures the overall trends in the linearized formulation, it does introduce errors as illustrated in the figure. 

\section{Turbomachinery case study}
In this section, the previously derived monotonic density-matching formulation is applied to fan stage OUU. The motivation for this case study comes from the work of \cite{Zamboni}, where the authors investigate the effects of variations in rear-seal leakage flows on fan stage performance. 

\subsection{Computational methods}
\begin{figure}[t!]
\centering
\includegraphics[natwidth=993,natheight=354,width=18cm]{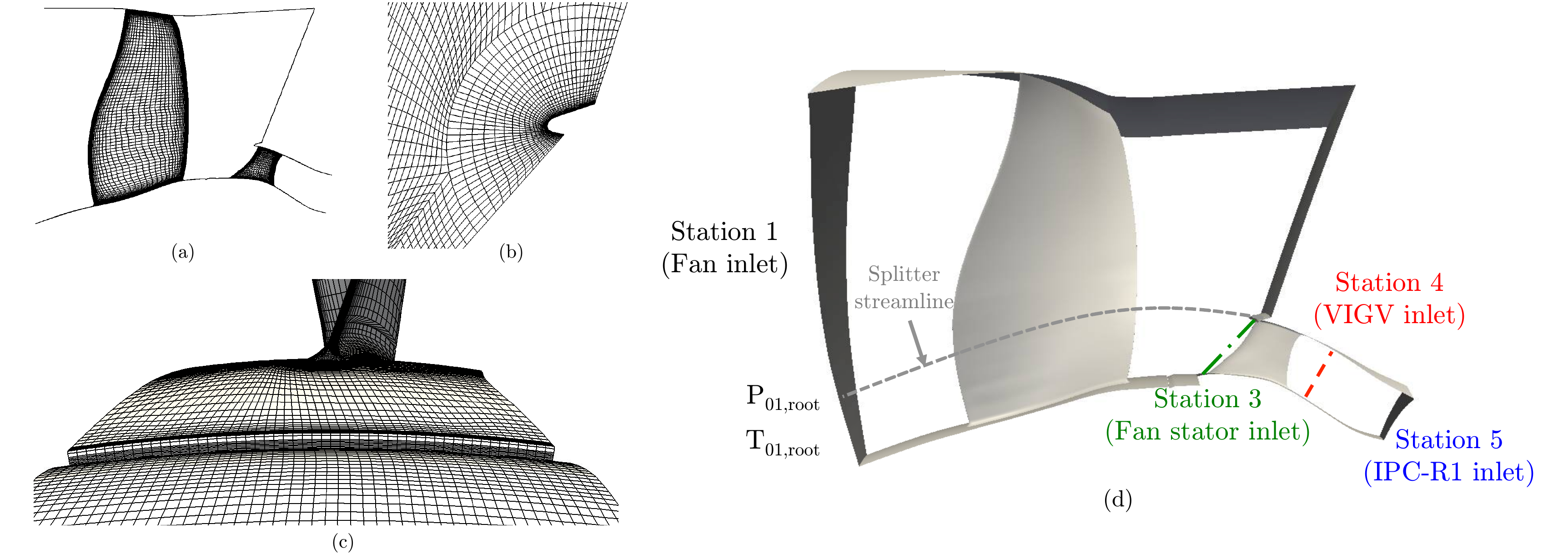}
\caption{Mesh for numerical computations: (a) Fan rotor and stator computational domains; (b) Close-up of the splitter C-mesh; (c) Downstream fan rotor cavity. (Note: in these images not all the cells are shown for clarity); (d) Fan stage computational domain with labeled stations (Note: images have been scaled for proprietary reasons)}
\label{mesh}
\end{figure}

For the subsequent computational investigations, the PADRAM-HYDRA\cite{Shahrokh,Lapworth} suite of codes is used. These have been previously validated in the authors' work in \cite{Seshadri_LEAK}. The computational domain considered in this work comprises of the fan stage of a modern aero-engine, comprising of the fan rotor, fan splitter, fan stator and the intermediate pressure (IP) compressor duct. The fan zone is modeled in the rotating frame, while the stator is in the stationary frame. A meridional view of the computational domain is shown in Figure~\ref{mesh}. The rotor zone mesh, which encompasses the bypass and core exits, the splitter, the inlet, the rotor blade and the hub, has 2.682 million cells (see Figure~\ref{mesh}(a). It uses a H-O-H grid topology with an H-mesh for the tip gap. The downstream stator also uses the same grid H-O-H grid topology and has 1.6 million cells. To capture the splitter fore of the fan stator a C mesh was utilized as shown in (b). The figure in (c) shows the cavity located between the rotor and the stator zones (as seen from the stator). It is between this patch that the rear-seal leakage flow emanates. This patch incorporates a platform radius edge, which permits control over the leakage inlet whirl angle. In (d) the entire computational domain is shown with the various stations. It should be noted that although the full fan stage is modeled, the primary domain of interest extends strictly from the fan hub to the splitter streamline. The inlet and outlet of this domain are given by Stations 1 and 5 respectively.

Nonlinear flow solves were carried out using HYDRA with a Spalart-Allmaras turbulence closure. A subsonic inflow boundary condition was utilized where 2D profiles of the inlet total pressure, total temperature and the Spalart variable (for turbulence) were prescribed. Both the stator exit and the bypass used a subsonic exit capacity boundary condition. The interface between the rotating and stationary zones was modeled with a mixing plane boundary condition that computes mass-average values at the rotor-core exit and feeds them to the stator inlet boundary condition. The fan rotor cavity was prescribed with a subsonic inlet whirl mass-flow boundary condition. Convergence was achieved after approximately 900 iterations using 4 multi-grid steps. Adjoint flow solves were also carried out using HYDRA with Spalart-Allmaras and with wall functions. The adjoint code is based on the discrete adjoint. This approach involves linearizing the set of discrete Navier-Stokes equations. In HYDRA the adjoint equations are solved with a time-marching method that employs partial updates of the numerical smoothness and viscous fluxes with Jacobi preconditioning at select stages in the Runge-Kutta iteration \cite{Giles}. 

\subsection{Design space, objectives and targets}
\begin{figure}[t!]
\centering
\includegraphics[natwidth=962,natheight=362,width=13cm]{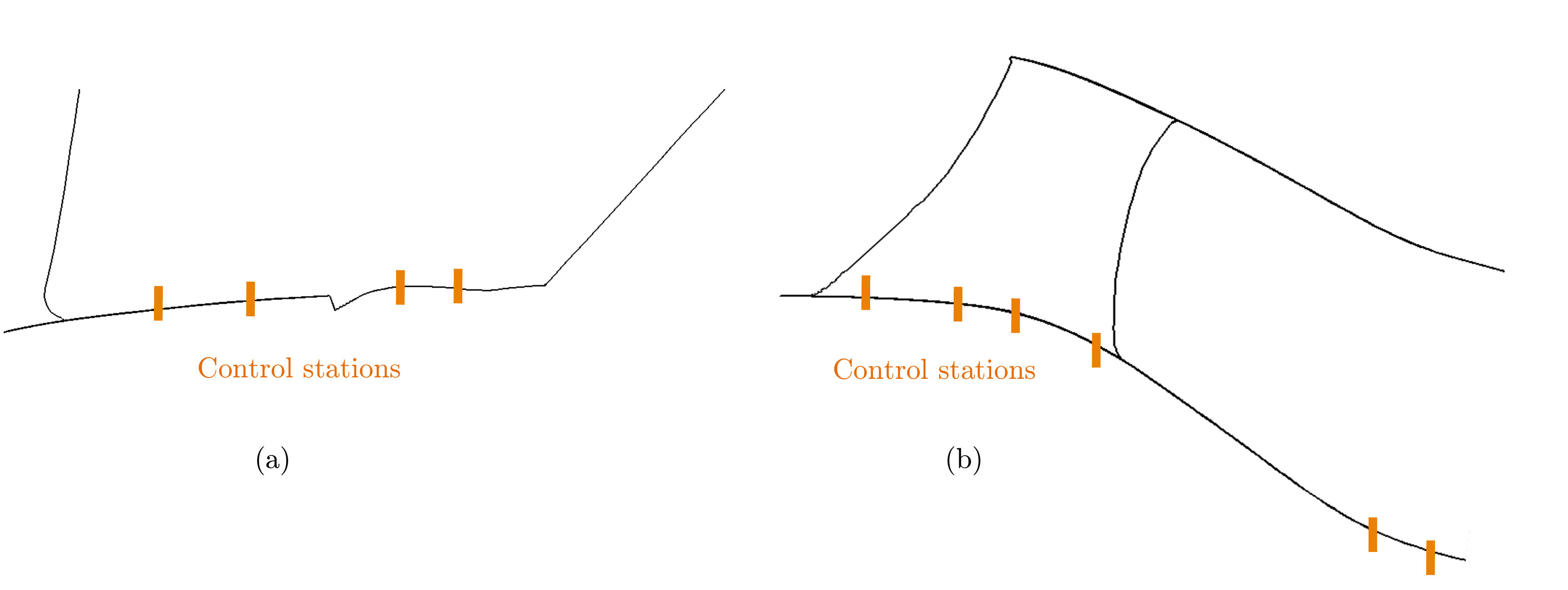}
\caption{Control point locations used for endwall perturbation via univariate b-spline curves: (a) For the rotor hub; (b) For the stator hub.}
\label{fandesign}
\end{figure}

A total of 35 design parameters were used to parameterize the fan stage geometry. Specifically, the focus of interest was the fan rotor downstream hub line, the fan stator hub line and the stator 3D design. The fan rotor geometry was not altered. The stator 3D parameterization is based on 3D blade engineering parameters. This includes tangential lean, skew, leading and trailing edge re-cambering and axial sweep (see \cite{Gallimore} for definitions of these degrees of freedom). The aforementioned five degrees of freedom are prescribed at 5 radial stations corresponding to 0,25,50,75,100 percent span, and then linearly interpolated to create the resulting blade profile. This yielded a total of 25 design variables for the 3D stator design. The endwall design perturbations used were based on a radial displacement of a chosen control point, that is then smoothly merged with the undisplaced points on the endwall geometry via a series of univariate b-spline curves. A total of four control stations were chosen immediately aft of the fan root, adjacent to the cavity, and a total of six control stations were chosen along the stator endwall (see Figure~\ref{fandesign}). For the latter, four stations were placed in-line with the stator, and two points were placed downstream. It should be noted that all six perturbations were axisymmetric. 

\begin{figure}[t!]
\centering
\includegraphics[natwidth=869,natheight=442,width=15cm]{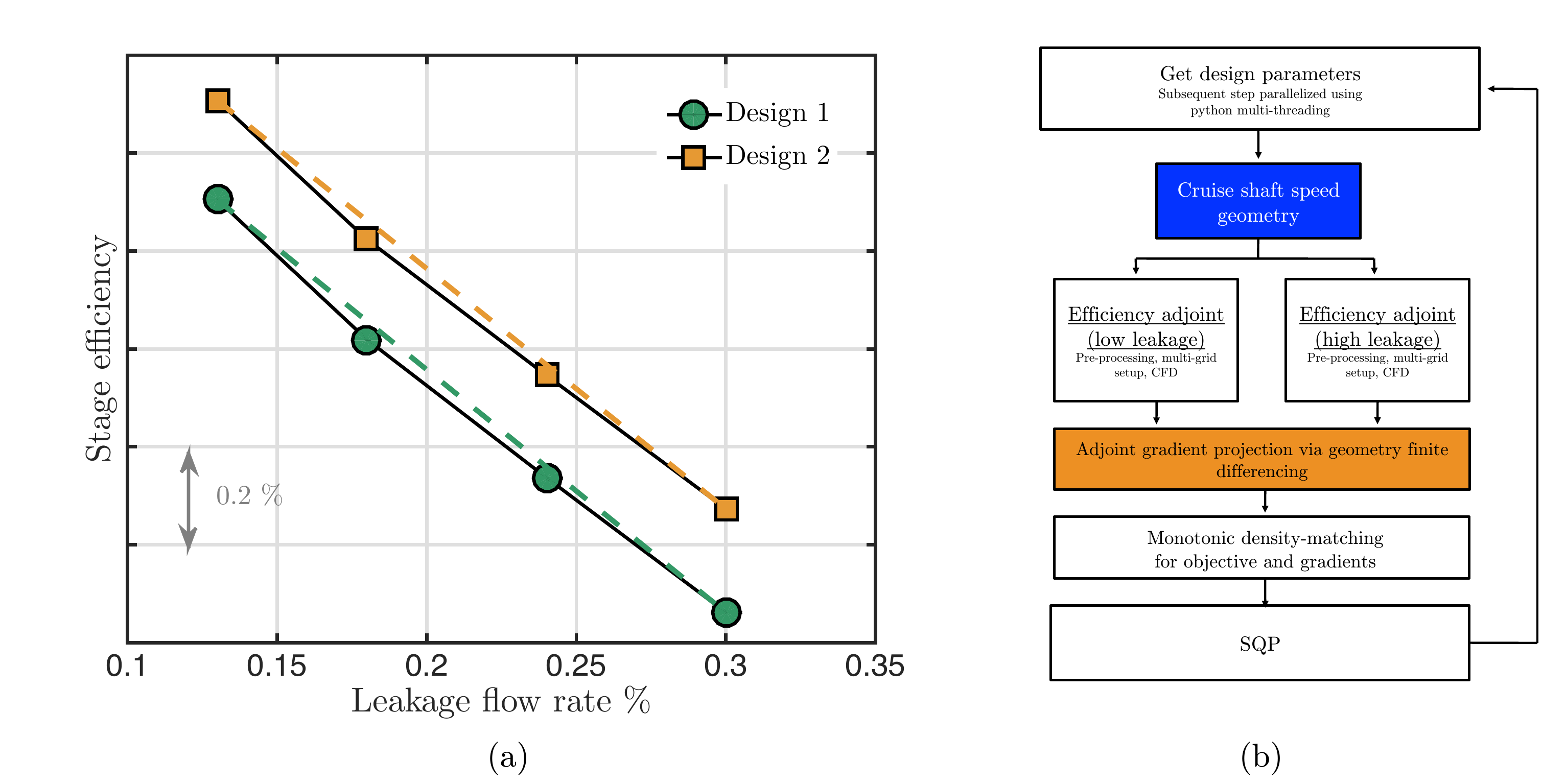}
\caption{(a) Monotonic relation between efficiency and rear seal leakage rate; (b) Optimization workflow}
\label{uncons_workflow}
\end{figure}

The qoi for this study was fan root efficiency, which is given by $\eta_{root}=\left(PR_{root}^{\left(\gamma-1\right)/\gamma}-1\right)/\left(TR_{root}-1\right)$, where $\gamma$ is the gas constant, and $PR_{root}=\left(P_{05}/P_{01,root}\right)$ and $TR_{root}=\left(T_{05}/T_{01,root}\right)$ are the pressure and temperature ratios respectively. These are computed from the inlet (see hub portion of Station 1 in Figure~\ref{mesh}(d)) and outlet stations (Station 5). To test the monotonic approach, an unconstrained optimization is pursued. The resulting workflow is shown in Figure~\ref{uncons_workflow}(b). For each vector of design variables, two nonlinear and adjoint flow solves are run in parallel corresponding to the minimum (0.13$\%$) and maximum ($0.30\%$) rear seal leakage mass-flow conditions. Once the nonlinear and adjoint flow solves are completed, the distance and its gradients are computed using the monotonic formulation detailed above. Distance values in all the cases were normalized by the value of the first function call. Here, individual nonlinear and adjoint flow solves were carried out on 48 cores and took approximately half a day. In total $48 \times 6 = 288$ cores were used. The sequential quadratic program (SQP) optimizer within the Smart Optimization for Turbomachinery (SOFT\cite{SOFT}) toolkit was used to carry out the optimization. The convergence criterion was a maximum of 40 function calls with a tolerance in the design parameters of $10^{-5}$. 

A beta(1.7,2.8) uncertainty in the rear seal leakage mass-flow rate was assumed as it closely resembled the expected variation over engine life. Three target exchange distributions were selected and are shown in Figure~\ref{optimized}. While all three targets had a lower variance, targets A and B had a higher mean, while in target C the mean was reduced. 

\subsection{Monotonic density-matching results}
\begin{figure}[t!]
\centering
\includegraphics[natwidth=1172, natheight=643, width=15cm]{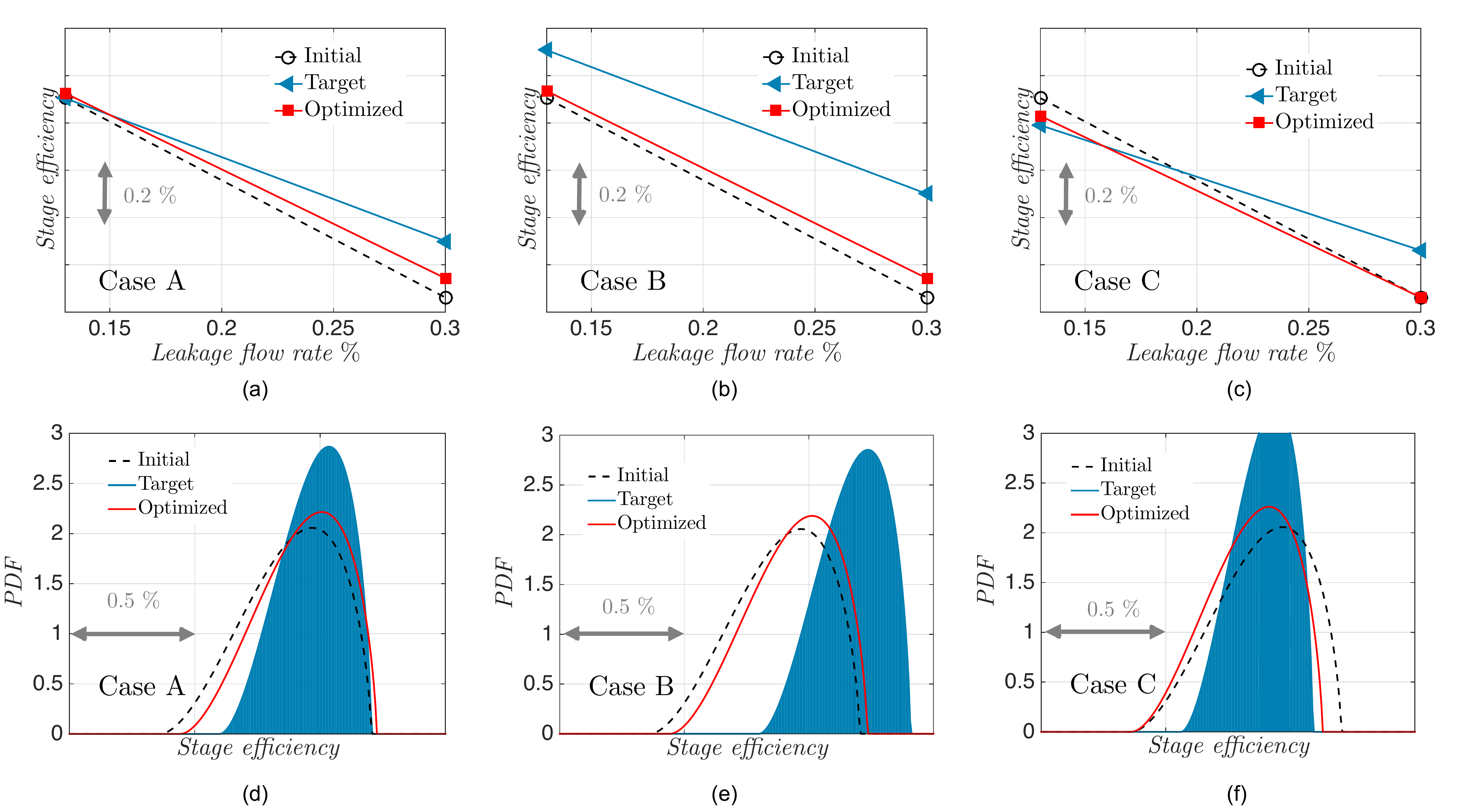}
\caption{Initial design, target design and optimized design exchange curves for cases A, B and C in (a); (b); (c); with corresponding pdfs in (d); (e); (f)}
\label{optimized}
\end{figure}

The results for the three optimizations are shown in Figure~\ref{optimized}. In terms of number of evaluations, Case A required 36 function calls (2 nonlinear flow solves and 2 adjoint calls per function call) with 6 major SQP iterations, Case B required 23 function calls with 4 major iterations and Case C had 37 function calls with 6 major iterations. Relative to the initial design, which is the outcome of numerous industrial optimization techniques and is thus already highly efficient, Case A reported a 0.04$\%$ increase in efficiency with a variance of 0.0262; Case B reported a 0.05$\%$ increase with a variance of 0.0368; Case C had a decrease in efficiency of 0.05$\%$ with a variance of 0.0252. The variance for the initial design was 0.0304. 

\begin{figure}[t!]
\centering
\includegraphics[natwidth=906, natheight=326, width=15cm]{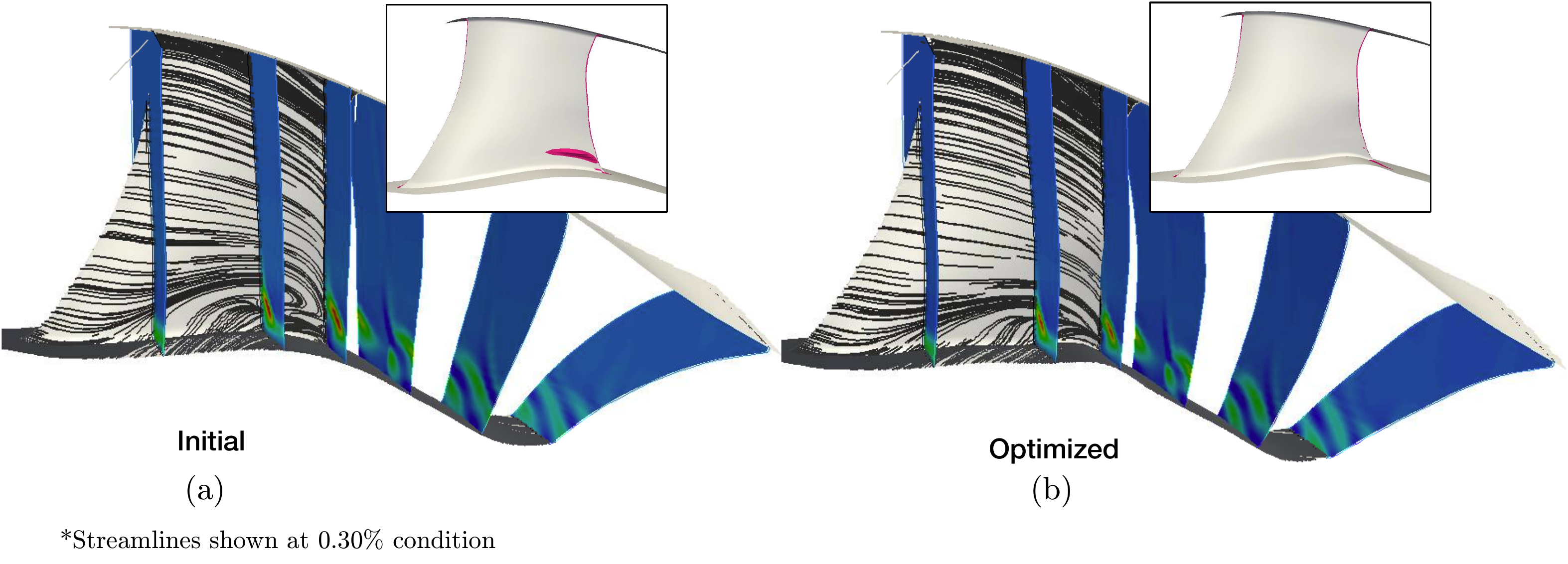}
\caption{Comparison of initial and optimized streamlines with snapshots of the negative axial velocity iso-volume for the (a) initial design; (b) optimized design A}
\label{aerofinal}
\end{figure}

In all three cases, the optimizer aimed to get as close to the target as possible. While the increases in efficiency are meagre, there is a clear effect on the aerodynamics. For instance, for Case A a significant reduction in the suction side stator separation was found as shown in Figure~\ref{aerofinal}, which lead to a smaller variance as the flow did not separate at the higher leakage flow rates. In Case A this was achieved largely by a change in the stator endwall curvature. Similar results with slightly different design solutions were found in Cases B and C. A more thorough treatment of the `robust aerodynamics' obtained through the techniques illustrated in this paper will be featured in a forthcoming publication.

\section{Conclusions}
In this paper an efficient approach to OUU was presented specifically for monotonic models. This new approach is a non-kernel density estimate variant of density-matching and uses the derived distribution theorem to exactly determine the output qoi’s pdf. The technique is successfully demonstrated on a real turbomachinery test case.  

\section*{Acknowledgements}
This research was funded through a Dorothy Hodgkin Postgraduate Award, which is jointly sponsored by the Engineering and Physical Sciences Research Council (EPSRC) (UK) and Rolls-Royce plc. The authors are grateful to David Radford and Mark Wilson for their assistance in various parts of this project. 

\bibliographystyle{asmems4}
\bibliography{references}
\end{document}